# Self-sovereign Identity – Opportunities and Challenges for the Digital Revolution

(Preprint, English Translation)

Uwe Der[1], Stefan Jähnichen[2,3], Jan Sürmeli[2,3]

**Abstract.** The interconnectedness of people, services and devices is a defining aspect of the digital revolution, and, secure digital identities are an important prerequisite for secure and legally compliant information exchange. Existing approaches to realize a secure identity management focus on central providers of identities such as national authorities or online service providers. Hence, changing residence or service provider often means to start over and creating new identities, because procedures for data portability are missing. *Self-sovereign digital identities* are instead created and managed by individuals, and enable them to maintain their digital identities independent from residence, national eID infrastructure and market-dominating service providers.

## Table of Contents

## 1. Digital Identities without Barriers

Digital identities are part of everyday life – not only in Europe but all over the world: Internet-based services require secure identification of their users – traditionally via user names and passwords, but also via special software and hardware such as chip cards, security tokens or the electronic ID (eID) of an ID card. In addition to the identification of people, digital identities of non-human entities such as companies or technical devices for the exchange of data and values are also playing an increasingly important role, as it is necessary to ensure that the data originates from the claimed source. Economic developments such as *Smart Manufacturing*, *Connected Car* or the *Internet of Things* show that the trustworthy identification of persons, systems or sensors is the key to the acceptance of such services.

A *digital identity* is a snapshot of the actual identity of a person, a company, a device, a car – more generally: an entity. The actual identity encompasses all the determining characteristics of an entity, which makes an entity distinguishable from others. Each digital identity consists of only a fragment of the identity and is usually created for a specific purpose in a specific context – to use a particular service or to interact with another entity. The individual digital identities differ in their level of detail: With respect to the supplied properties (such as name, age, etc.), the accuracy of their description and the degree of abstraction (for example: complete name vs. name on identity card plus birth name, date of birth/precise age vs. age of legal majority). Furthermore, a digital identity has a clear temporal point of reference: Characteristics of an entity can change at least partially (e. g. address, bank account details, weight, health status, stage of illness, highest school leaving certificate or the maintenance status of a non-human entity). Each human, each system, each sensor, i. e. each entity, therefore has only one identity but an unlimited number of digital identities - a number that can quickly go into the hundreds for a human being.

---

[1] IT Service Omikron GmbH, Berlin. uwe.der@itso-berlin.de

[2] Technische Universität Berlin, Berlin. stefan.jaehnichen@tu-berlin.de, jan.suermeli@tu-berlin.de

[3] FZI Forschungszentrum Informatik am Karlsruher Institut für Technologie, Karlsruhe. stefan.jaehnichen@fzi.de, jan.suermeli@fzi.de

The term *secure digital identities* integrates the requirements of *privacy* and *trustworthiness*: Privacy means that only authorized persons, institutions or systems can access a digital identity (and the information contained therein), where the authorization, eg. to access certain data, is granted by the entity described by the digital identity. Trustworthiness refers to the correctness of the information contained in the digital identity, i. e. that the characteristics contained therein actually belong to the entity with a certain level of assurance.

The following three examples illustrate the global scope of secure digital identities:

The public-private partnership ID2020 aims at supplying every human being with an access to a personal, persistent, private and portable digital identity (ID2020, 2017). The initiative is primarily dedicated to individuals who have so far been denied such access – a problem affecting more than one billion people worldwide. ID2020 follows the UN Sustainable Development Programme, which aims to create a "legal digital identity" for every human being in the world by 2030 (United Nations, 2015).

In 2015, more than 244 Million people lived in a different country than they were born in (United Nations, 2016). While the reasons for migration differ, the inherent requirements for identity management are similar for different cases: Those who temporarily or permanently change their place of residence need to retain access to their "former" identity: health data collected in the previous place of residence are a prerequisite for further treatment plans, certificates and proof of qualification are prerequisites for employment in the new place of residence. In addition to migrants who permanently change their place of residence, these requirements also apply to students who complete part of their studies abroad or to people who have several places of residence for professional reasons.

The EU is stepping up its efforts to create or strengthen a European digital single market. Cross-border digital identities are an essential key to success, as are common legal requirements such as the eIDAS regulation (EUR-Lex, 2014) und the new EU General Data Protection Regulation (GDPR) (EUR-Lex, 2016), which aim to remove economic and organizational barriers between the EU member states, and to motivate citizens and companies to act more strongly on the European internal market.

While these examples relate to human beings and their digital identities, current developments such as the *Internet of Things* (IoT) show that such requirements also hold for non-human entities such as companies, devices and cars. The high degree of decentralized architectures of interconnected devices requires secure digital identities which are available on demand independent of their place of operation: The Smart Home should also be accessible from different spots on the world, Connected Cars should be able to cross borders, and a wearable measuring health data should not stop recording when in foreign countries.

This global scope points to a potential limitation of national or otherwise geographically limited administrative approaches. In general, there are two approaches to tackling this problem: either the globalization of administrative bodies (e. g. through bilateral and multilateral agreements), or their reduction to a minimum in favor of self-sovereignty. Today, the first method is common practice: responsibility is placed in the hands of globally active institutions from the business world, which, while offering comfort, often act against the interests of their customers for economic reasons by accepting the invasion of privacy and acting in a non-transparent manner. Self-sovereign digital identities pose the opposite approach: responsibility for administration is transferred to the entity itself. The opportunities and challenges of this new approach to the management of digital identities of individuals are discussed in this article and an implementation of the approach in the form of the ISÆN concept is described. Finally, the possibility of extending the concepts for human beings to other entities such as devices or institutions is considered.

## 2. Self-Sovereign Digital Identities for Persons

Decentralization is a visible trend in identity management in order to tackle the issues of globally scoped digital identities. (Allen, 2016). In the beginning of the world wide web, IP-addresses were centrally assigned by the Internet Assigned Numbers Authority. This form of Centralized Identity was superseded by the concept of Federated Identity: Microsoft Passport enabled the usage of multiple internet services with a single account. Other providers founded the Liberty Alliance. The next step in the road to decentralization was the emerge of User-centric Identity, pushed by initiatives such as OpenID, focusing more on the control of personal data by the user, and stressing that identity management should be implemented independent from service usage. However, the majority of online identities is still created by single, powerful service providers such as Facebook or Google. As a result, the identity is still controlled by a company, in contrast to the user. For instance, the service provider could always revoke the identity, and thus also for connected services, without the user being able to intervene.

*Self-sovereign identities* could be the next step: Every person creates and manages their own digital identities. Following (Allen, 2016), self-sovereign identity can be characterized by the following ten commandments:

- existence of the identity of a person independent of identity administrators or providers,
- the person being in control of their digital identities,
- the person having full access to their own data,
- systems and algorithms are transparent,
- digital identities are persistent,
- digital identities being portable,
- digital identities being interoperable,
- data economy being enforced, and
- the rights of the person being protected.

Self-sovereign identities give the person more control over their digital appearance, but the person now is responsible for the measures taken to establish and maintain both privacy and trustworthiness. Since the digital identities are not issued by third parties, trustworthiness is achieved by the person obtaining evidence for the correctness of the information contained in the digital identity from third parties. Such evidence has then to be provided by the person if necessary. For example, an address contained in a digital identity can be confirmed by a registration office and, if necessary, the person can submit this confirmation to a service provider later. Some identification documents, such as the "neue Personalausweis" (nPA, new German ID card), for example, already offer suitable technologies, and there exist also services that offer automatic verification of ID documents (Keesing Technologies, 2017).

This also shifts the problem of trustworthiness: instead of proving the authenticity of the information, the person must prove the authenticity of the evidence. This can be achieved by third parties signing the documents in order to prove its authenticity. The issuer of the receipts does not take on the role of an intermediary: the collection of receipts is only loosely linked to the actual transaction and can be carried out independently of this. (Rannenberg, Camenisch, & Sabouri, 2015; Verifiable Claims Working Group, 2017). Since certain characteristics of a person, such as address or account details, can change, these evidences are potentially only valid for a limited period of time. It is also conceivable that a person may request documents for authenticity from different third parties and use them as needed. In turn, the recipient of the personal data, i.e. the service provider can request a certain level of trustworthiness for a transaction, for example "confirmed by a public authority in the EU", and the person can then select appropriate documents corresponding to the required level of assurance.

## Opportunities of Self-Sovereign Identities

Following the ten commandments of self-sovereign identity, a person ultimately has full control over their own digital identities regardless of the current living conditions, including where the person currently resides, the person's citizenship or the chosen service providers for email or social networks. Self-sovereign identity therefore increases the freedom of the individual and could also counteract the oligopoly structure of today's Internet, in which the management of most digital identities is the responsibility of the "Big Five" (Apple, Microsoft, Google, Amazon, Facebook).

The ten commandments of self-governed identity already cover some of the criteria demanded by ID2020, namely the requirements of personality, persistence, privacy and portability. If the trustworthiness of digital identities is no longer directly tied to local governments, migrants and people with multiple domiciles will retain their identity even after relocations. This independence could also have positive impact on the situation of persons living in crisis areas or in areas affected by corruption.

The new EU General Data Protection Regulation (GDPR, valid from May 2018) strengthens a person's rights by requiring increased transparency, data economy and data portability. It also calls for the implementation of concepts such as privacy by default and privacy by design, envisioning that future solutions always take the highest possible data protection settings as default, and understand the protection of privacy as an important design objective. Self-sovereign identity not only gives individuals the highest level of control over its data, but selective sharing of personal data with service providers follows the idea of data economy and privacy by default/design. The transparency created by self-governance could also strengthen the European digital single market by removing barriers of missing trust. For companies, self-sovereign identity can offer new ways to fulfil their duties to provide useful information to their customers.

## Challenges of Self-Sovereign Identities

If a person uses an identity provider such as Facebook Connect, the person delegates all management to that service. This includes the responsibility to protect their privacy and to ensure the trustworthiness of their data. However, the person typically has little insight into the actual handling of his digital identities. At the same time, the person has to make only few decisions, as eg. about the technology for the secure storage or transmission of data. The person thus exchanges some of his control over his digital identities for a certain comfort. The core challenge is therefore to offer solutions that help persons to manage the additional administrative efforts, i. e. to make the use of self-sovereign identity not only possible but also sufficiently comfortable.

This includes the generation and storage of a person's multiple digital identities as well as the management of evidences for the trustworthiness of these digital identities. In addition, methods are required for performing and logging transactions of rights of use to other persons and services. Problems arise in particular if trusted third parties are to be avoided as much as possible:

- Protection of the privacy of persons for transactions of usage rights, in particular also more complex problems such as the difficulty of prohibiting profiling by third parties,
- Transparency between the persons and services involved in the transaction, in particular, ensuring consensus about the content and conduct of the transaction,
- Persistency of digital identities and logs of transactions, to ensure long-term transparency,
- Trustworthiness of digital identities and hence pieces of evidence,
- Consistency between the usage rights granted in a transaction and the real usage of the data,
- Data formats and standardized interfaces for securely exchanging evidences and digital identities.

Although these problems are not all new, and probably not even characteristic for self-sovereign identities, this new identity management approach requires the development of new solutions or at least the adaptation of current ones. For some of these challenges, the use of Distributed Ledger technologies such as Blockchains (Nakamoto, 2008; Pilkington, 2016) and Smart Contracts (Meitinger, 2017) are proposed to log and possibly even implement both the verification of digital identities and the transaction of personal data or usage rights (Jacobovitz, 2016; Civic, 2017; Humaniq, 2017; ævatar.coop, 2017; Jähnichen, et al., 2017).

At first glance, digital identities exist independently of jurisdictions. However, the identification of the legal framework applicable to a transaction is not trivial on the Internet, since the question of "where" a transaction takes place is difficult to answer. To this end, the GDPR has introduced the principle of *Lex loci solutionis*, which shifts its scope from the actual location of the transaction to the citizenship of the persons involved. Overall, it is necessary to support individuals and service providers in recognising, understanding the applicable judicial area and assisting them to behave lawfully in this scope.

## 3. ISÆN: Standardizing Self-Sovereign Identities for Human Beings

The standardization initiative CEN Workshop 84 on a 'Self-Sovereign Identifier (s) for Personal Data Ownership and Usage Control' (CEN WS ISÆN, 2016) proposes the ISÆN concept for managing digital identities of human beings. The goal is an overall concept for self-sovereign identities that is in line with the EU General Data Protection Regulation (GDPR). The standard is intended to cover the following areas in particular:

- Creation of a core identity by the human,
- derivation of transaction-based digital identities from the human's core identity,
- implementation for requesting and granting explicit consent from the human, and
- logging all transactions in a public distributed ledger.

The core identity is formed from the characteristics of the human and contains not only identifying features such as the name but also biometric features such as a fingerprint or face recognition data. Identities are bound to single transactions and formed with one-way functions from the core identity. This allows a human to prove their identity without outsiders being able to establish links between different transactions. Logging transactions in a tamper-proof memory enables humans to gain an overview of who processes which of their data. At the same time, companies can also make use of the transaction protocols to fulfil their disclosure obligations, and the tamper-proof nature of the logging process gives them additional legal certainty.

The concept was evaluated in a study for the German Federal Ministry for Economic Affairs and Energy of Germany (Bundesminsterium für Wirtschaft und Energie) and its further development was recommended (Jähnichen, et al., 2017).

## 4. Outlook: Self-Sovereign Identities for Things and Institutions

The Internet of Things is the strong networking of clearly identifiable "things", such as everyday objects, production machines, products, networked cars, without central administrative instances. Typical identification methods of such "things" are RFID chips or barcodes referring to a digital identity. As with individuals, decentralized management of digital identities is also of interest.

Smart contracts enable automated processing of business logic via public distributed ledgers such as Ethereum (Ethereum Foundation, 2017). Secure, digital identities are the key to preventing fraudulent execution of smart contracts. In addition to people and things, the digital identities of companies and other organisations also play a central role.

While there are already approaches for self-sovereign identity for human beings, the extension of the concept to non-human entities raises new questions: How could self-sovereign identity be conceived and implemented for non-human entities? How can a non-human entity recognize and characterize its own identity? What features of a non-human identity are as unique and difficult to imitate as biometric features in humans?

## References

Allen, C. (2016, April 27). *The Path to Self-Sovereign Identities*. Retrieved October 31, 2017, from Coindesk: https://www.coindesk.com/path-self-sovereign-identity/

CEN WS ISÆN. (2016). *Self-Sovereign Identifier(s) for Personal Data Ownership and Usage Control*. Retrieved October 31, 2017, from https://www.cen.eu/work/areas/ICT/Pages/WS-IS%C3%86N.aspx

Civic. (2017). *Civic.com*. Retrieved from https://www.civic.com

Ethereum Foundation. (2017). *ethereum - Blockchain App Platform*. Retrieved from https://www.ethereum.org/

EUR-Lex. (2014, July 23). *Verordnung (EU) Nr. 910/2014 des Europäischen Parlaments und des Rates vom 23. Juli 2014 über elektronische Identifizierung und Vertrauensdienste für elektronische Transaktionen im Binnenmarkt und zur Aufhebung der Richtlinie 1999/93/EG*. Retrieved October 31, 2017, from http://eur-lex.europa.eu/legal-content/DE/ALL/?uri=CELEX:32014R0910

EUR-Lex. (2016, April 4). *Verordnung (EU) 2016/679 des Europäischen Parlaments und des Rates vom 27. April 2016 zum Schutz natürlicher Personen bei der Verarbeitung personenbezogener Daten, zum freien Datenverkehr und zur Aufhebung der Richtlinie 95/46/EG (Datenschutz-Grundverordnung)*. Retrieved October 31, 2017, from http://eur-lex.europa.eu/legal-content/DE/TXT/?uri=CELEX:32016R0679

ævatar.coop. (2017). *aevatar.com*. Retrieved from https://aevatar.com

Humaniq. (2017). *Humaniq.com*. Retrieved from https://humaniq.com

ID2020. (2017). *An Alliance Committed to Improving Lives Through Digital Identity*. Retrieved from http://id2020.org/

Jähnichen, S., Weinhardt, C., Müller-Quade, J., Huber, M., Rödder, N., Karlin, D., . . . Shaar, P. (2017). *Sicheres Identitätsmanagement im Internet: Eine Analyse des ISÆN-Konzepts (Individual perSonal data Auditable addrEss) durch die Smart-Data-Begleitforschung im Auftrag des Bundesministeriums für Wirtschaft und Energie*. http://www.digitale-technologien.de/DT/Redaktion/DE/Downloads/Publikation/smartdata_studie_isaen.html.

Jacobovitz, O. (2016). *Blockchain for Identity Management.* The Lynne and William Frankel Center for Computer Science Department of Computer Science. Beer Sheva, Israel: Ben-Gurion University.

Keesing Technologies. (2017). *AuthentiScan; Professional Authentication. Straight forward solution.* Retrieved October 31, 2017, from https://www.keesingtechnologies.com/automated-id-checking/

Meitinger, T. H. (2017). Smart Contracts. *Informatik-Spektrum*(40), 371-375.

Nakamoto, S. (2008). *Bitcoin: A Peer-to-Peer Electronic Cash System*. Retrieved 2017, from http://bitcoin.org/bitcoin.pdf

Pilkington, M. (2016). Blockchain Technology: Principles and Applications. In F. X. Olleros, & M. Zhegu (Eds.), *Handbook of Research on Digital Transformations*.

Rannenberg, K., Camenisch, J., & Sabouri, A. (Eds.). (2015). *Attribute-based Credentials for Trust: Identity in the Information Society*. Springer.

United Nations. (2015). *UN Sustainable Development Goals*. Retrieved October 31, 2017, from http://www.un.org/sustainabledevelopment/

United Nations. (2016, January 2016). *244 million international migrants living abroad worldwide, new UN statistics reveal*. Retrieved October 31, 2017, from UN Sustainable Goals: http://www.un.org/sustainabledevelopment/blog/2016/01/244-million-international-migrants-living-abroad-worldwide-new-un-statistics-reveal/

Verifiable Claims Working Group. (2017). *About the Verifiable Claims Working Group*. Retrieved from https://www.w3.org/2017/vc/WG/